\magnification=\magstep2
\def\para{\par\noindent}
\def\sqr#1#2{{\vcenter{\vbox{\hrule height.#2pt
        \hbox{\vrule width.#2pt height#1pt \kern#1pt
          \vrule width.#2pt}
        \hrule height.#2pt}}}}

\newcount\notenumber

\def\note{\advance\notenumber by 1
\footnote{$^{\the\notenumber}$}}
\baselineskip 20pt
\centerline{{\bf Persistence in the Zero-Temperature Dynamics of the}}
\centerline{{\bf Diluted Ising Ferromagnet in Two Dimensions}}\para
\vskip 0.25cm
 \para S.~Jain,
 \para School of Mathematics and Computing,\para
 University of Derby,\para
 Kedleston Road,\para
 Derby DE22 1GB,\para
 U.K.\para
\vskip 0.25cm
\para E-mail: S.Jain@derby.ac.uk 
\vskip 3.0cm
\para Classification Numbers:
\para 05.20-y, 05.50+q, 05.70.Ln, 64.60.Cn, 75.10.Hk,
 75.40.Mg 
\vskip 1.5cm
\para 
\para Published in Physical Review E60, R2445 (1999)
\vskip 1.0cm
\para Cond-mat/9906272
\para
\vfill\eject
\para {\bf ABSTRACT}
\para The non-equilibrium dynamics of the strongly diluted random-bond
Ising model in two-dimensions $(2d)$ is investigated numerically.
 The persistence probability, $P(t)$, of spins
which do not flip by time $t$ is found to decay to a
 non-zero, dilution-dependent, value $P(\infty)$. We find 
that $p(t)=P(t)-P(\infty)$ decays
 {\it exponentially} to zero at large times. Furthermore, the fraction of
spins which {\it never} flip is a monotonically increasing function
over the range of bond-dilution considered. Our findings,
 which are consistent with a recent result of Newman and
 Stein, suggest that persistence in diluted and pure systems falls into
  different
classes. Furthermore, its behaviour would also appear to depend crucially 
on the strength of the dilution present.  
\para 
\vfill\eject
\para In the non-equilibrium dynamics of spin systems at zero-temperature we
are interested in the fraction of spins, $P(t)$, that persist in the same state
up to some later time $t$.
 For homogeneous ferromagnetic Ising models in $d$-dimensions, $P(t)$, has
been found to decay algebraically [1-4] 
$$P(t)\sim t^{-\theta (d)},\eqno(1)$$
for $d < 4$, where $\theta (d)$ is the new non-trivial persistence exponent.

\para The presence of a non-vanishing $P(t)$ as $t\to\infty$ has been reported 
in computer simulations of both the Ising model in higher dimensions 
($d > 4$) [3]
 and the $q-$state Potts model in $2d$ for $q > 4$ [5]; this feature is
 sometimes referred to as \lq blocking\rq . Obviously, if
$P(\infty) > 0 $, we can reformulate the problem by restricting our attention
only to those spins that eventually flip. Hence, we can consider
 the behaviour of
$$p(t)=P(t)-P(\infty).\eqno(2)$$

\para Although the numerical simulations of the $q-$state Potts
 model mentioned above [5] seem to
indicate that $p(t)$ also decays algebraically, the evidence is by no means 
conclusive.

\para By considering the dynamics of the local order parameter the
 persistence problem can be generalised to
 non-zero temperatures [6-9].

It is only recently [10,11] that attention has turned to the persistence 
problem in systems containing disorder. Numerical simulations of the
 zero-temperature dynamics of the weakly-diluted
 Ising model in $2d$ [10] also
reported that $P(\infty) > 0$. In fact, the study in [10] is consistent with
the presence of three distinct regimes: an initial short time 
regime where the behaviour is pure-like; an intermediate regime where the
persistence probability decays logarithmically; and a final long
 time regime where the system \lq freezes\rq\ and $P(t)$ is effectively
 constant. 

\para Very recently, Newman and Stein [11] have argued that the
 \lq blocking\rq\ [5]
 of spins in systems with continuous disorder
 is associated with the fact that \lq {\it every spin flips only finitely
 many times}\rq\ . As a consequence, in some simple $1d$ models 
 $p(t)$ was found to decay exponentially rather
than algebraically for large times, namely
$$p(t)\sim e^{-kt},\eqno(3)$$
where $k>0$. 
 In contrast, persistence in the weakly-diluted Ising model appears to decay
 logarithmically in the 
intermediate regime. (Note that [10] examines the behaviour of $P(t)$ and
 not $p(t)$). 

\para Clearly, it is of immense interest to establish whether the presence of
\lq blocking\rq\ in a system necessarily implies exponential decay of the 
persistence probability. Howard [12] has found evidence for exponential decay
in certain non-disordered models with \lq blocking\rq\ ($2d$ hexagonal lattices
and Bethe lattices with $z=3$ are discussed in [12]).

 To clarify and further investigate the situation,
 in this Rapid Communication we present the results
 of computer simulations
of an Ising model containing {\it strong} bond 
dilution. Here we restrict our attention to zero-temperature.

The Hamiltonian of the model we work with is given by
$${\it H} = -\sum_{<ij>} {J_{ij} S_i S_j}\eqno(4)$$
where $S_i=\pm 1$ are Ising spins situated on every site of a
 square $L\times L (=N)$ lattice with periodic boundary conditions
 and the summation runs
 over all nearest-neighbour pairs only. The 
quenched ferromagnetic exchange interactions are selected from
 a binary distribution given by
$$P(J_{ij}) = (1-p)\delta(J_{ij}) + p \delta(J_{ij}-1)\eqno(5)$$
where $p$ is the concentration of bonds.

  We obtained data for $L=500$ and $750$ at zero temperature for a broad
range of bond-concentrations 
 ($0\le p\le 0.5$) on 
a suite of Silicon
 Graphics workstations and for $L=1000$ on a SGI Origin 2000; as the data
for the different lattice sizes studied are practically indistinguishable,
 here we
 simply
present the results for the largest lattice simulated.
\para 
We begin each run with a random starting configuration of the spins and
then update the lattice by first calculating the energy change that would
result from flipping a spin. The rule we use is: always flip if the energy
 change is negative, never flip if the energy change is positive and flip at
random if the energy change is zero.  
\para  The number, $n(t)$,
of spins which have never flipped until time $t$ is then counted. As we are
working with strongly diluted lattices, it is necessary to monitor the
 value of $n(t)$ after practically each Monte Carlo step.

\para The persistence 
probability is given by [1] 
$$P(t)=[<n(t)>]/N\eqno(6)$$
where $<\dots>$ indicates an average over different 
initial conditions and $[\dots]$ denotes an average over samples i.e. over
 the bond-dilution. For the simulations considered in this work we averaged
over at least 100 different initial conditions and samples for each run.

 We now discuss our results. To examine the decay of the persistence
probability, in Fig. 1 we plot $\ln P(t)$ versus
$\ln t$ for a wide range of bond concentrations, $p: 0.1\le p\le 0.5$, for
a lattice of size $L=1000$. The 
decay of $P(t)$ appears to be non-algebraic before \lq freezing\rq\ occurs. 
We see that, effectively, $P(t)=P(\infty)$ for $t> t^*(p)$, where the value
$t^*(p)$ depends on the strength of the dilution. 
Furthermore, the non-zero value of $P(\infty)$ also depends on
 $p$, with the fraction of non-flipping spins increasing 
monotonically with the bond concentration. The increase in $P(\infty)$ with
$p$ can be seen more clearly in Fig. 2 where we have plotted some additional
data at values of the exchange interaction not shown in Fig. 1. The numerical
values of $P(\infty)$ for the different bond concentrations simulated are
also displayed in Table 1.

Obviously, when $p=0$ all spins eventually flip as the energy change in
 flipping is always zero.
 For a value of $p\ne 0$, there will be regions of the lattice containing
 finite clusters where it will cost energy to
flip spins. For example, an isolated bond connecting two up spins is just such
a stable cluster.
The occurrence of these clusters
increases with the bond concentration and hence also does 
the fraction of spins which never flip. This increases smoothly to $p=0.5$, the
bond percolation threshold, where it appears to level off. That is, the maximum
value of $P(\infty)\sim 0.46$. Clearly, $P(\infty)$ must
decrease eventually for higher values of $p$ as we know that every spin flips
infinitely many times for
the pure model, $p=1$ [1, 11].  

 We now consider the non-algebraic decay of $P(t)$ to $P(\infty)$.
As discussed earlier, it is more convenient to work with $p(t)$ from Eqn. (2).
In Fig. 3 we replot the data displayed in Fig. 1 as $\ln p(t)$ against $t$.
 The straight
 lines are
linear fits to Eqn. (3) after discarding data for short times. 
It is evident from Fig. 3
 that $p(t)$ indeed decays exponentially at large times. Hence, we confirm that
for the strongly diluted Ising model in $2d$ persistence
 decays exponentially as predicted by Newman and Stein [11].
 This is in marked contrast to the behaviour for the pure
[1-3] and the weakly diluted models [10].    

 To conclude, we have presented new data for the zero-temperature
 dynamics of the
strongly diluted random-bond $2d$ Ising ferromagnet. This system exhibits \lq
blocking\rq\ and we
find evidence that $p(t)$ decreases exponentially 
for large times. The fraction of spins which {\it never} flip increases
monotonically from zero with increasing bond concentration. Our results support
the suggestion that the decay of the persistence probability can be
 non-algebraic for certain classes of models. Indeed, for the diluted $2d$
Ising model the behaviour of $p(t)$ would appear to depend crucially on
the strength of the dilution.
\para {\bf Acknowledgement}
\para I am grateful to C.M. Newman and D.L. Stein for commenting on the draft
version of this paper.
 I would like to acknowledge Matthew Birkin 
 for both technical assistance and maintaining the Silicon
Graphics workstations. The CPU time on the SGI Origin 2000 at the University
of Manchester was made available by the Engineering and Physical Sciences
Research Council (EPSRC), Great Britain.
\vfill\eject 
\para FIGURE CAPTIONS
\vskip 1cm
\para Fig. 1
\para Log-log plot of $P(t)$ versus $t$ for the bond-diluted $2d$ Ising
model for a range of bond concentrations, $p$; the size of the lattice is
$1000\times 1000$.
\vskip 1cm
\para Fig. 2
\para A plot of the fraction of spin which NEVER flip ($P(\infty)$)
 against the bond concentraion $p$.
\vskip 1cm
\para Fig. 3
\para Plot of $\ln p(t)$ against $t$ for different bond concentrations, $p$.
The straight lines are linear fits to the data after discarding the initial
short time behaviour.
\vfill\eject
\para TABLES
\vskip 1cm
\para Table 1
\vskip 0.5cm
\vbox{\tabskip=0pt \offinterlineskip
\def\tablerule{\noalign{\hrule}}
\halign to250pt{\strut#& \vrule#\tabskip=1em plus2em&
\hfil#& \vrule#& 
\hfil#& \vrule#\tabskip=0pt\cr\tablerule
&&\hidewidth $p$\hidewidth&&
\hidewidth $P(\infty)$\hidewidth&\cr\tablerule
&& 0   && 0        &\cr\tablerule
&& 0.025&& 0.0708(1)&\cr\tablerule
&& 0.050&& 0.1340(1)&\cr\tablerule
&& 0.075&& 0.1900(1)&\cr\tablerule
&& 0.100&& 0.2390(1)&\cr\tablerule
&& 0.125&& 0.2813(1)&\cr\tablerule
&& 0.150&& 0.3173(1)&\cr\tablerule
&& 0.175&& 0.3479(2)&\cr\tablerule
&& 0.200&& 0.3732(2)&\cr\tablerule
&& 0.250&& 0.4097(1)&\cr\tablerule
&& 0.300&& 0.4331(2)&\cr\tablerule
&& 0.350&& 0.4453(3)&\cr\tablerule
&& 0.400&& 0.4526(3)&\cr\tablerule
&& 0.450&& 0.4559(2)&\cr\tablerule
&& 0.500&& 0.4576(1)&\cr\tablerule
\noalign{\smallskip}}}
\vskip 1.0cm
\para The fraction of spins, $P(\infty)$, which {\it never} flip at various
values of the bond concentraion, $p$.
\vfill\eject
\para REFERENCES
\item {[1]} B. Derrida, A. J. Bray and C. Godreche, J.Phys. A {\bf 27},
 L357 (1994).
\item {[2]} A.J. Bray, B. Derrida and C. Godreche, Europhys. Lett. {\bf 27},
 177 (1994).
\item {[3]} D. Stauffer J.Phys.A {\bf 27}, 5029 (1994).
\item {[4]} B. Derrida, V. Hakim and V. Pasquier, Phys. Rev. lett. {\bf 75},
 751 (1995); J. Stat. Phys. {\bf 85}, 763 (1996).
\item {[5]} B. Derrida, P.M.C. de Oliveira and D. Stauffer, Physica {\bf 224A},
604 (1996).
\item {[6]} S. N. Majumdar, A. J. Bray, S. J. Cornell, and C. Sire, Phys. Rev.
 Lett. {\bf 77}, 3704 (1996).
\item {[7]} K. Oerding, S. J. Cornell, and A. J. Bray, Phys. Rev. E{\bf 56}, R25
(1997).
\item {[8]} B. Zheng, Int. J. Mod. Phys. B{\bf 12}, 1419 (1998).
\item {[9]} J-M. Drouffe and C. Godreche, e-print cond-mat/9808153.
\item {[10]} S. Jain, Phys. Rev. E{\bf 59}, R2496 (1999).
\item {[11]} C.M. Newman and D.L. Stein, Phys. Rev. Lett. {\bf 82}, 3944 (1999).
\item {[12]} C.D. Howard, preprint (1999).
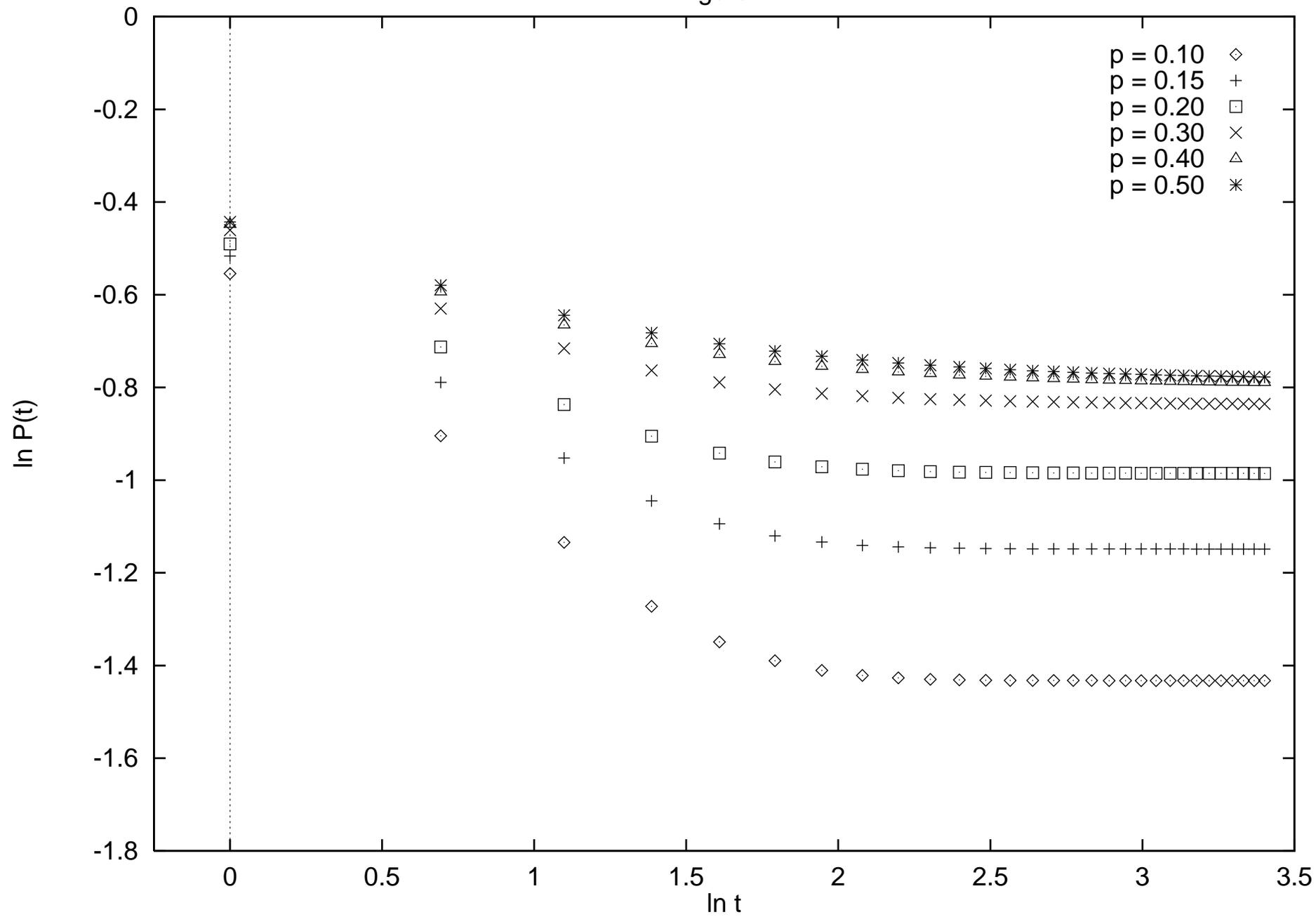
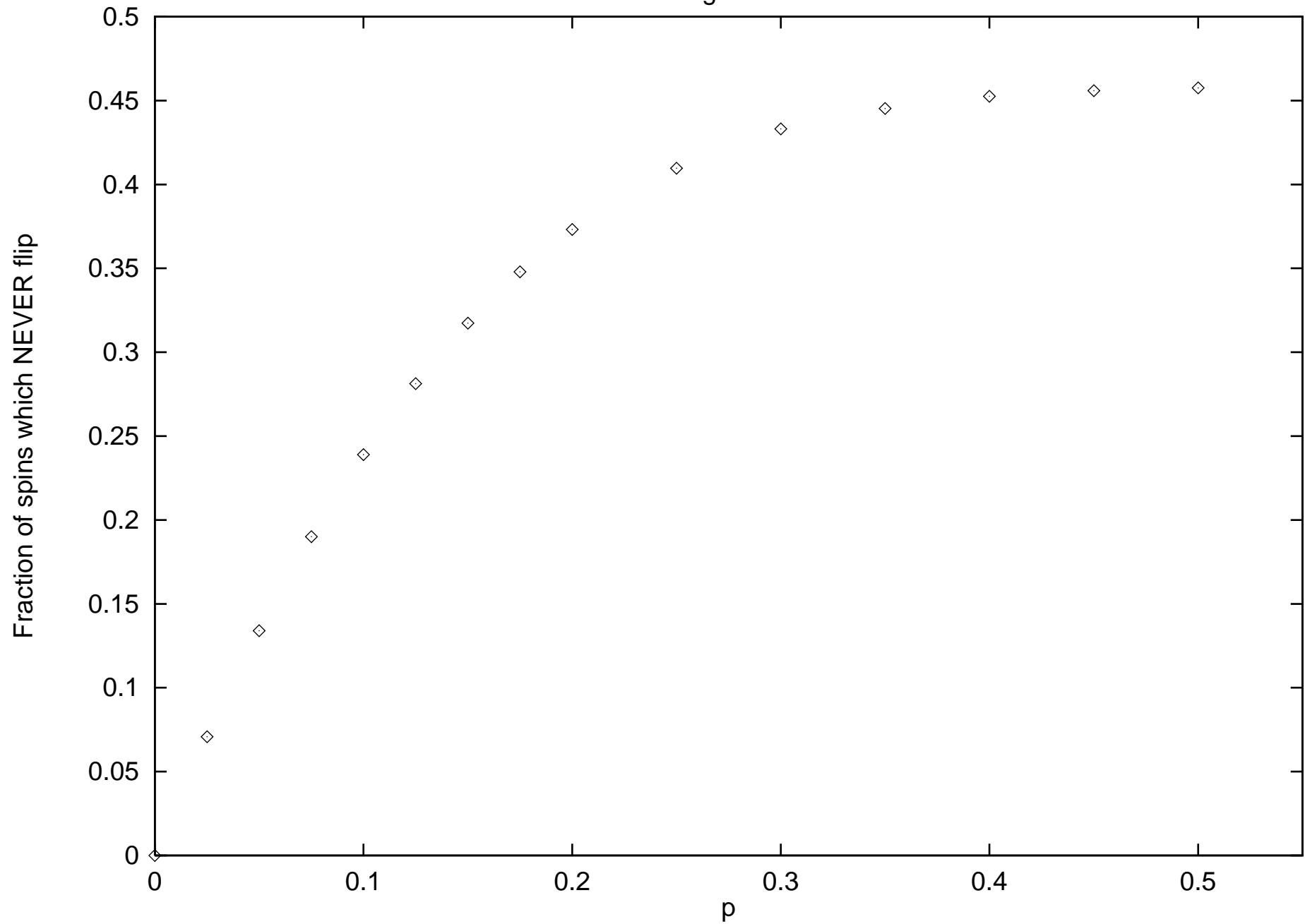
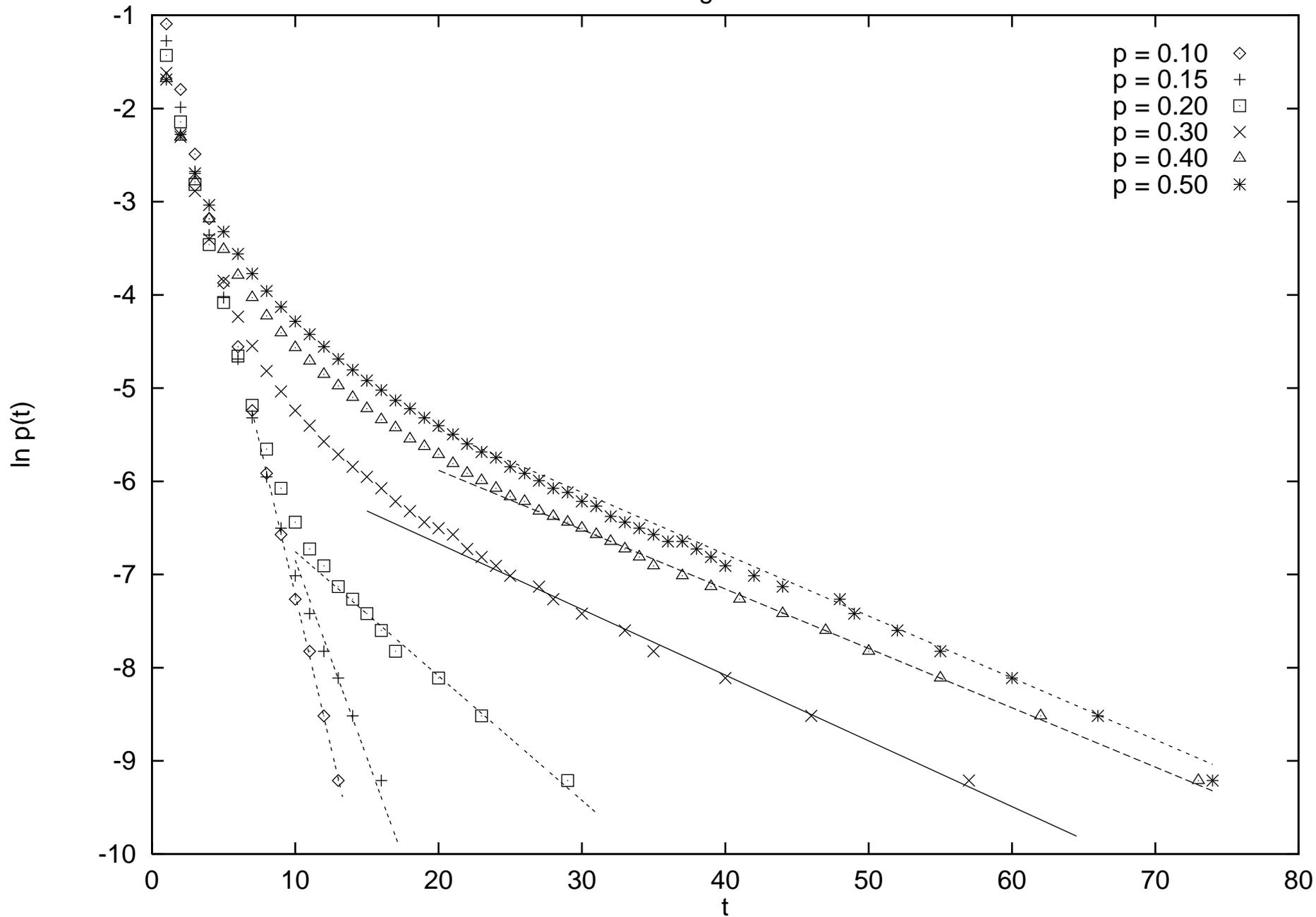
\end